\documentclass[letterpaper,twocolumn,prl,aps,superscriptaddress,amsmath,amssymb,floatfix]{revtex4-1}
\usepackage{mathptmx}
\usepackage[latin9]{inputenc}
\usepackage{color}
\usepackage{amsmath}
\usepackage{amssymb}
\usepackage{graphicx}
\usepackage{esint}
\usepackage[unicode=true,
 bookmarks=true,bookmarksnumbered=false,bookmarksopen=false,
 breaklinks=false,pdfborder={0 0 1},backref=false,colorlinks=true]
 {hyperref}
\hypersetup{
 linkcolor=magenta,urlcolor=blue,citecolor=blue,pdfstartview={FitH},hyperfootnotes=false}

\makeatletter

\usepackage{textcomp}
\usepackage{epstopdf}

\usepackage{amsfonts}

\pdfpageheight\paperheight
\pdfpagewidth\paperwidth

\@ifundefined{textcolor}{}{%
 \definecolor{BLACK}{gray}{0}
 \definecolor{WHITE}{gray}{1}
 \definecolor{RED}{rgb}{1,0,0}
 \definecolor{GREEN}{rgb}{0,1,0}
 \definecolor{BLUE}{rgb}{0,0,1}
 \definecolor{CYAN}{cmyk}{1,0,0,0}
 \definecolor{MAGENTA}{cmyk}{0,1,0,0}
 \definecolor{YELLOW}{cmyk}{0,0,1,0}
}

\usepackage{xcolor}\usepackage{soul}
\definecolor{blue}{rgb}{0,0,1}
\definecolor{red}{rgb}{1,0,0}
\definecolor{green}{rgb}{0,1,0}

\makeatother

\begin{document}
\title{Classical-to-quantum transition in multimode nonlinear systems with strong photon-photon coupling}
\author{Yue-Xun~Huang}
\affiliation{CAS Key Laboratory of Quantum Information, University of Science and Technology of China, Hefei 230026, P. R. China.}
\affiliation{CAS Center For Excellence in Quantum Information and Quantum Physics,
	University of Science and Technology of China, Hefei, Anhui 230026, P. R. China.}
\author{Ming~Li}
\email{lmwin@ustc.edu.cn}

\affiliation{CAS Key Laboratory of Quantum Information, University of Science and Technology of China, Hefei 230026, P. R. China.}
\affiliation{CAS Center For Excellence in Quantum Information and Quantum Physics,
	University of Science and Technology of China, Hefei, Anhui 230026, P. R. China.}
\author{Ke~Lin}
\affiliation{School of Software, Tsinghua University, Beijing, 100084, China}
\author{Yan-Lei~Zhang}
\affiliation{CAS Key Laboratory of Quantum Information, University of Science and Technology of China, Hefei 230026, P. R. China.}
\affiliation{CAS Center For Excellence in Quantum Information and Quantum Physics,
	University of Science and Technology of China, Hefei, Anhui 230026, P. R. China.}
\author{Guang-Can~Guo}
\affiliation{CAS Key Laboratory of Quantum Information, University of Science and Technology of China, Hefei 230026, P. R. China.}
\affiliation{CAS Center For Excellence in Quantum Information and Quantum Physics,
	University of Science and Technology of China, Hefei, Anhui 230026, P. R. China.}
\author{Chang-Ling~Zou}
\email{clzou321@ustc.edu.cn}

\affiliation{CAS Key Laboratory of Quantum Information, University of Science and Technology of China, Hefei 230026, P. R. China.}
\affiliation{CAS Center For Excellence in Quantum Information and Quantum Physics,
	University of Science and Technology of China, Hefei, Anhui 230026, P. R. China.}

\begin{abstract}
	With advanced micro- and nano-photonic structures, the vacuum photon-photon coupling rate is anticipated to approach the intrinsic loss rate and lead to unconventional quantum effects. Here, we investigate the classical-to-quantum transition of such photonic nonlinear systems using the quantum cluster-expansion method, which addresses the computational challenge in tracking large photon number states of the fundamental and harmonic optical fields involved in the second harmonic generation process. Compared to the mean-field approximation used in weak coupling limit, the quantum cluster-expansion method solves multimode dynamics efficiently and reveals the quantum behaviors of optical parametric oscillations around the threshold. This work presents a universal tool to study quantum dynamics of multimode systems and explore the nonlinear photonic devices for continuous-variable quantum information processing.
\end{abstract}
\maketitle

\section{Introduction}

Nonlinear optics has been exploited in abundant classical and quantum optics applications since the advent of lasers~\cite{boyd2003nonlinear, agrawal2000nonlinear}. Under the current theoretical framework, the simplest approach to describe a coherent optical field is by characterizing it with only one parameter, i.e., the field amplitude, with the system dynamics governed by a set of nonlinearly coupled equations among modes of different amplitudes. This treatment, known as the mean-field approximation (MFA)~\cite{walls2007quantum},  however neglects
the influence of the quantum fluctuations. In a more rigorous framework, the optical fields are treated as Gaussian states for which the mean-field amplitude and second-order correlations are assumed to be complete to describe the system~\cite{wang2007quantum, weedbrook12}. Then, the quantum
fluctuations and correlations of optical fields can be derived by
calculating the covariance matrix. 
In the past decades, the MFA and Gaussian-state approximation have been
widely applied in quantum optics and successfully predict a variant
of phenomena, including the squeezing~\cite{walls1983squeezed},
continuous-variable entanglement~\cite{braunstein2005quantum}, and
the thermal dynamics of mechanical resonators~\cite{PhysRevLett.98.030405,PhysRevA.84.032317}.

In conventional nonlinear optics systems, the nonlinear coupling strengths
between optical modes are much weaker than the dissipation rates,
therefore the higher-order correlations between modes are negligible
due to the strong decoherence, and the Gaussian state approximation
is accurate.
As the fabrication technique and material improves, the photon-photon coupling in the nonlinear system can be greatly enhanced, to the point that  the single-photon nonlinearity becomes appreciable in photonic integrated circuits. The coupling strength to dissipation rate ratio $g/\kappa$ has been greatly boosted during the last decades~\cite{Krastanov2021}, using a microresonator made by gallium arsenide (GaAs)~\cite{chang2019strong}, aluminum nitride (AlN) \cite{Bruch17000}, indium gallium phosphide (InGaP)$\:$\cite{zhao2021nanophotonic}, lithium niobate (LN)~\cite{Lu:19,chengya,chen2021photon} and etc. For example, Lu et al. demonstrated a $g/\kappa$ ratio over $1\:\%$ in a periodically poled lithium niobate microring resonator \cite{lu2020toward}, suggesting significant nonlinear effects at the level of tens photons. At this high nonlinearity  limit, the Gaussian state approximation no longer holds. With even larger $g/\kappa$, the nonlinear system is predicted to exhibit atom-like features~\cite{PhysRevLett.96.057405,li2020photon}, thus significant quantum effects arise under excitation at the single-photon level~\cite{KerrBlock,majumdar2013single}. It is intriguing to explore the classical-to-quantum transition in this new regime where $g/\kappa$ approaches unity and the conventional treatment of high amplitude bosonic modes under MFA is no longer valid, and the quantum master equation with truncated Fock-state dimension becomes inefficient.

In this Letter, the classical-to-quantum transition of nonlinear $\chi^{(2)}$ processes is investigated based on the quantum cluster-expansion (QCE) approach. In particular, we focus on degenerate $\chi^{(2)}$ interactions, including second-harmonic generation and optical parametric oscillation at different nonlinear coupling rates and pump powers. The numerical results show the deviation of the mean photon numbers and the quantum statistics from the predictions by the classical theory, and manifest the classical-to-quantum transition when increasing the $g/\kappa$ and the pump power. We developed the code for generating recursive QCE to arbitrary orders with an arbitrary number of modes. By comparing with the conventional numerical approaches based on master equations of truncated Hilbert space, the validity of QCE is verified and shows a $10^{4}$ times speedup under excitation of only $400$ photons. Our approach is efficient for solving the problems with large intracavity photon numbers and also moderate $g/\kappa$ ratio, and could be extended to study
the quantum behaviors of other complex nonlinear optics systems. 
\begin{figure}
	\begin{centering}
		\includegraphics[width=1\columnwidth]{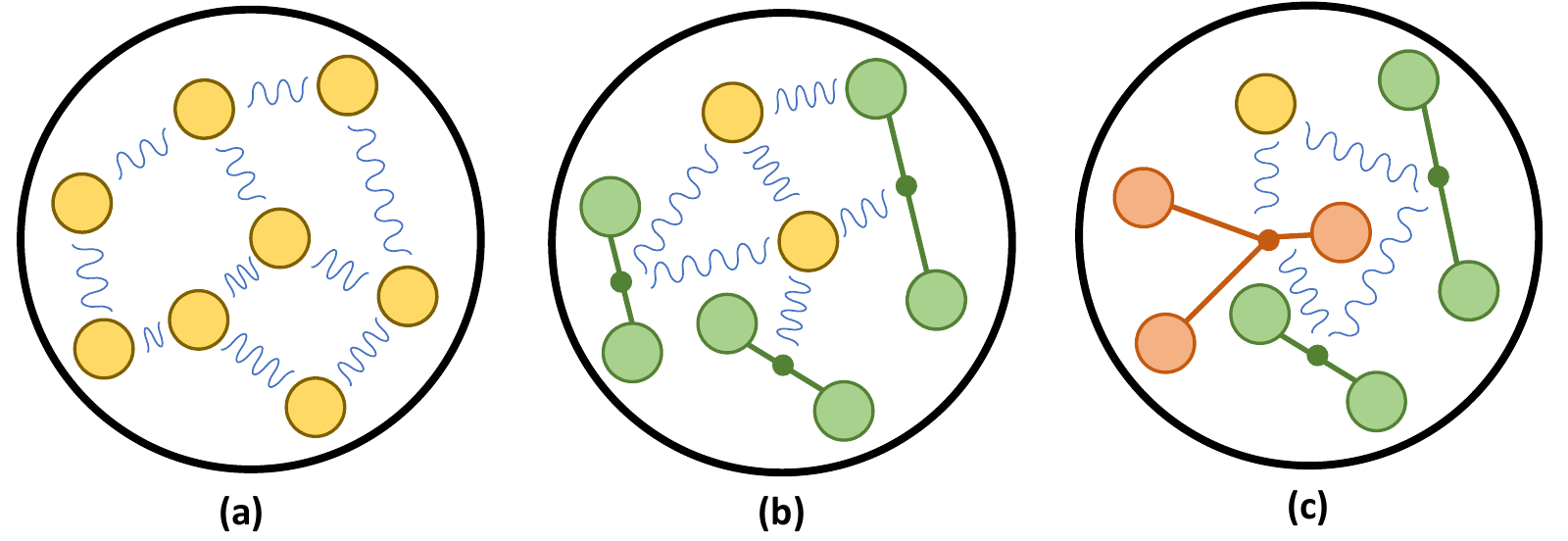}
		\par\end{centering}
	\caption{Illustration of a quantum nonlinear system under different orders
		of quantum cluster expansion (QCE), with circles linked by straight
		lines denoting a cluster of bosonic operators and wavy lines coupling between clusters. (a) 1st order QCE, corresponding
		to the mean-field approximation. (b) 2nd order QCE. (c) 3rd order
		QCE. For $M$-th order QCE, the system is described by the dynamics of operators with orders no higher than $M$.}
	\label{Fig1}
\end{figure}

\section{Principle of QCE}

In resonance-enhanced nonlinear photonics processes such as
three-wave mixing and frequency comb generation in  microrings~\cite{li2018enhancement}, there are multiple optical
resonances simultaneously satisfying the energy and phase matching conditions. These modes are generally described by the bosonic annihilation
operators $O_{j}$, where $j$ labels individual mode, and any given system operator could be written as the product of a cluster
of mode operators $A=\Pi_{j,k}O_{j}^{\dagger m_{j}}O_{k}^{n_{k}}$,
which is a $M$-th order operator with $M=\sum_{j}m_{j}+\sum_{k}n_{k}$.
For an open quantum system, the dynamics of an operator $A$ follows
the master equation~\cite{walls2007quantum}
\begin{eqnarray}
	\frac{d}{dt}\rho & = & -\frac{i}{\hbar}\left[H,\rho\right]+\sum_{j}\kappa_{j}\mathcal{L}_{d_{j}}[\rho],\label{eq:dynamics}
\end{eqnarray}
where $\rho$ is the density matrix and $H$ is the Hamiltonian of
the system, $\kappa_{j}$ is the dissipation rate and $\mathcal{L}_{d_{j}}[\rho]=2d_{j}\rho d_{j}^{\dagger}-d_{j}^{\dagger}d_{j}\rho-\rho d_{j}^{\dagger}d_{j}$
is the Lindblad operator for a jump operator $d_{j}$. For example, in the case of a reservoir with near-zero thermal excitation, the amplitude dissipation of individual modes is $\kappa_{j}\mathcal{L}_{d_{j}}[\rho]$. The expectation
value of an operator $\langle A\rangle=\mathrm{Tr}\left\{ A\rho\right\} $
also follows the master equation 
\begin{eqnarray}
	\frac{d}{dt}\langle A\rangle & = & \frac{i}{\hbar}\langle\left[H,A\right]\rangle+\sum_{j}\kappa_{j}\langle\mathcal{L}'_{d_{j}}[A]\rangle.\label{eq:expectdy}
\end{eqnarray}
where $\mathcal{L}'_{d_{j}}[A]=2d_{j}^{\dagger}Ad_{j}-d_{j}^{\dagger}d_{j}A-Ad_{j}^{\dagger}d_{j}$.

The master equation shows that the dynamics of the expectation value
of the $M$-th order cluster $A$ is directly coupled to operators
$\left[H,A\right]$ and also $\mathcal{L}_{d_{j}}[A]$. For a simple bilinear Hamiltonian that consists of only $2$nd-order clusters, the evaluations of $O_{j}$ only depends on
the first order clusters, thus the master equations of the system
become an array of closed-form linear equations of $O_{j}$ and $O_{j}^{\dagger}$, as shown by Fig.~\ref{Fig1}(a).
The expectations of all 2nd-order clusters then can be derived
based on this set of linear equations, which further gives the covariance matrix of
the system~\cite{walls2007quantum}. If all the environment modes
and the system initial states are Gaussian states, 
the expectation values of the higher-order clusters can be expressed in terms of 
the first- and second-order clusters {[}Fig.~\ref{Fig1}(b){]}, and
the system quantum state evolution can be completely described by
the covariance matrix. However, for the system that involves $\chi^{(2)}$
and higher-order nonlinear interactions, i.e. the $H$ consists of
3-rd or higher order clusters, the dynamics of 2nd-order clusters
should depend on the higher order clusters {[}Fig.~\ref{Fig1}(c){]},
and the evolution of all clusters in general can not be obtained
in a closed form. For example, for $H=\hbar a^{\dagger}a\left(b+b^{\dagger}\right)$,
the evolution of the expectation value of $a$ is governed by
\begin{equation}
	\frac{d\langle a\rangle}{dt}=-i\langle a(b+b^{\dagger})\rangle-\kappa_{a}\langle a\rangle,
\end{equation}
which requires the values of $\langle ab\rangle$ and $\langle ab^{\dagger}\rangle$.
Repeating the same process for $\left[H,\,A\right],\,A=\{ab,ab^{\dagger}\}$ 
leads to an infinite hierarchy that $\langle A\rangle$ is expressed by an
infinite set of higher-order operators.

Two approaches can be applied to address this divergence of 
high-order operators: (i) MFA, which neglects the fluctuations of strong fields and replaces them by complex numbers, thus reducing the order of clusters. In the classical
limit, all operators are replaced by complex numbers, resulting in
MFA shown in Fig.$\:$\ref{Fig1}(a). (ii) Fock space truncation (FST).
When the number of excitations in bosonic modes are restricted, the
quantum state could be represented in the Fock basis with a finite dimension, and then the master equation could be solved numerically. Although both approaches are widely adopted in quantum optics studies, they are not applicable to a system with moderate
nonlinearities and strong drives.

Therefore, we revert to solve the original master equation with the
QCE approach. In practice, it is unrealistic to track the expectation
values of an infinite set of operators to evaluate the expectation value
$\langle A\rangle$. Although the number of clusters involved for
a complete system dynamics generally diverges, the QCE can
be solved approximately or even analytically by truncating the order
of QCE, i.e. setting the high order clusters as zero. Such treatment
has been established in quantum chemistry~\cite{harris2020algebraic}
and semiconductor systems~\cite{kira2001exciton} as well as Bose-Einstein
condensates beyond the mean-field theory~\cite{vardi2001bose}.  
Different from these previous
studies, the order of the hierarchy of multimode nonlinear photonic system depends on the specific nonlinear processes involved. At large mode number and orders,  the expansion of the clusters increases drastically, and it is intractable
to directly write down all the equations for clusters and solve them
analytically.

The high-order correlation is directly related to the nonlinear coupling rate $g$.
In the weak-coupling limit $g\ll\kappa$, the correlation
between different operators can be neglected. The expectation values of $N$-th order operators can be directly factorized to the product of $1$st order operators by
\begin{eqnarray}
	\langle\widehat{N}\rangle & \approx & \prod_{j=1}^{N}\langle O_{j}\rangle,\label{eq:1order}
\end{eqnarray}
which is the main assumption of the MFA. At this limit,  the photonic modes are treated as harmonic oscillators and the optical fields can be approximated to coherent states, as has been adopted by most experiments. The system dynamics is described by a set of nonlinearly coupled equations only containing the expectation values of $1$st operators [Fig.$\:$\ref{Fig1}(a)], while the tiny quantum fluctuations around the mean-field $\langle O_{j}\rangle$ is neglected.  As the $g/\kappa$ ratio increases, the strong anharmonicity leads to the distortion of the quantum state from the coherent state, in which case the quantum correlation becomes significant so that the high-order correlation can not be directly factorized to the product of single-order operators represented by Eq.$\:$(\ref{eq:1order}). For example, a $2$nd order
operator can be written as $\langle\widehat{2}\rangle=\langle\widehat{1}\rangle\langle\widehat{1}\rangle+\Delta\langle\widehat{2}\rangle=\langle\widehat{2}\rangle_{s}+\Delta\langle\widehat{2}\rangle,$
where $\langle\widehat{2}\rangle_{s}$ represents the MFA approximation
by Eq.$\:($\ref{eq:1order}) and $\Delta\langle\widehat{2}\rangle$
indicates the purely correlated part. Likewise, the factorization
of a $N$th order cluster reads
\begin{align}
	\langle\widehat{N}\rangle = & \langle\widehat{N}\rangle_{s}+\langle\widehat{N}-2\rangle_{s}\Delta\langle\widehat{2}\rangle+\langle\widehat{N}-4\rangle_{s}\Delta\langle\widehat{2}\rangle\Delta\langle\widehat{2}\rangle\nonumber  +\ldots\Delta\langle\widehat{N}\rangle\nonumber \\
	= & \langle O_{n}\rangle\langle\widehat{N}-1\rangle+\Delta\langle O_{n}\widehat{1}\rangle\langle\widehat{N}-2\rangle\nonumber \\
	& +\Delta\langle O_{n}\widehat{2}\rangle\langle\widehat{N}-3\rangle+\ldots\Delta\langle\widehat{N}\rangle,\label{eq:Norder}
\end{align}
where each product term presents one factorization and is
summed over all indistinguishable combinations.
$\widehat{i}$ denotes all the possible $i$th order cluster within the $N$th order cluster. And $\widehat{N}-i$ represents
the factorization of the remaining $\left(N-i\right)$th order cluster.

Based on Eqs.$\:$(\ref{eq:expectdy}) and
(\ref{eq:Norder}), the dynamics of a nonlinear system can be implemented
following the $M$th order QCE:
\begin{enumerate}
	\item To get the expectation value of $\langle\widehat{N}\rangle$, submit
	$\langle\widehat{N}\rangle$ to Eq.$\:$(\ref{eq:expectdy}) and one
	gets its relation with $\langle\widehat{N}\rangle=\langle\left[H,N\right]\rangle$
	and operator $\langle\widehat{i}\rangle$ of other orders.
	\item Following the $M$th truncation that all $\Delta\langle\widehat{N}\rangle=0$
	for $N\geq M$, $\langle\widehat{N}\rangle$ and $\langle\widehat{i}\rangle$
	can be factorized according to Eq.$\:$(\ref{eq:Norder}).
	\item Repeat Steps 1 and 2 for any cluster that appears in Step 2 until
	no new clusters are generated. In this way, we arrive at a set of
	nonlinear coupled equations involving clusters up to $M$-th orders.
\end{enumerate}
We implement an open-source package to automatically with tree based symbol system to complete the above procedure. The technical details about the code are offered online and the codes are available in Ref.~$\:$\cite{code}.

\section{Classical-to-quantum transition of $\chi^{(2)}$ interaction}

We apply the QCE approach to investigate the classical-to-quantum transition
of a signature quantum nonlinear optical system --- degenerate
$\chi^{(2)}$ interaction --- involved in the second-harmonic generation
(SHG) and optical parametric oscillation (OPO).
For the phase-matched degenerate $\chi^{(2)}$ interaction between
modes $a$ and $b$, the Hamiltonian can be written as~\cite{Guo2016,li2018enhancement}
\begin{eqnarray}
	H & = & \omega_{a}a^{\dagger}a+\omega_{b}b^{\dagger}b+g\left(a^{\dagger2}b+a^{2}b^{\dagger}\right),
\end{eqnarray}
where $a$ ($b$) and $a^{\dagger}$ ($b^{\dagger}$) are the annihilation
and creation operator of the fundamental (second-harmonic) mode with
$\omega_{a(b)}$ being the corresponding resonant frequency. $g$
is the nonlinear coupling strength that depends on the material and
cavity geometry.

\begin{figure}[t]
	\begin{centering}
		\includegraphics[width=1\columnwidth]{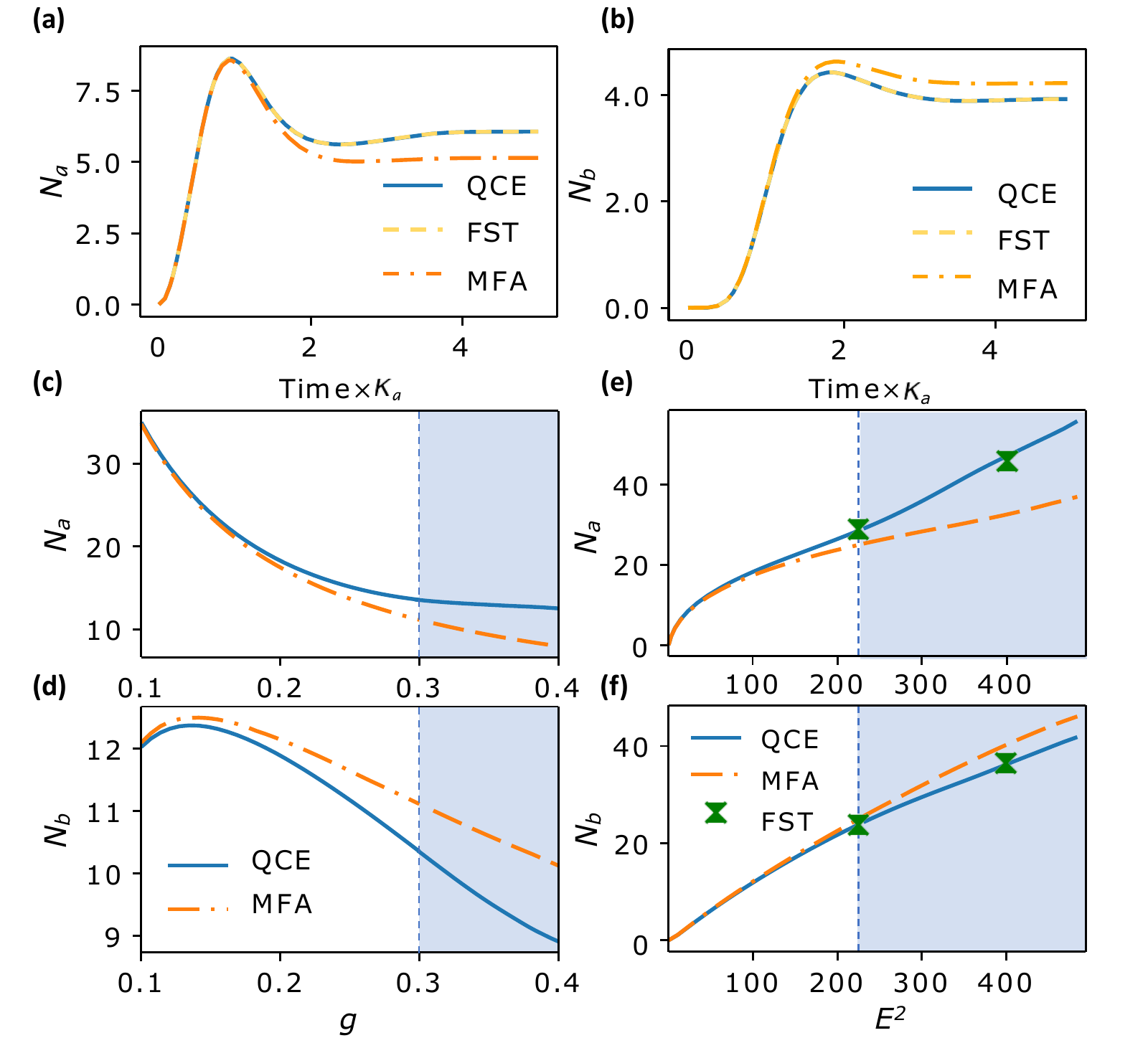}
		\par\end{centering}
	\centering{}\caption{Quantum-to-classical transition behaviors of 
		second-harmonic generation.
		(a, b) The dynamical evolution of intracavity photon numbers, with
		$E=6$ and $g=0.4$. (c, d) Steady-state intracavity photon number
		vs the coupling strength $g$, with $E=10$. (e, f) Steady-state
		intracavity photon number vs the external drive strength $E$,
		with $g=0.2$. In all calculations, $\kappa_{a}=\kappa_{b}=1$ and the evolution duration of the system is $T_{ss}=10/\kappa_{a}$ which is long enough for the system to reach the steady state. Solid
		lines, dashed lines and dash-dotted lines corresponds to 
		results obtained via 4th-order quantum-cluster expansion (QCE), Fock-space
		truncation (FST),
		and mean-field approximation (MFA), respectively. The shadow region
		denotes parameter spaces that lead to self-pulsing behavior
		predicted by MFA.}
	\label{Fig2}
\end{figure}

\begin{figure}[t]
	\centering{}\includegraphics[width=1\columnwidth]{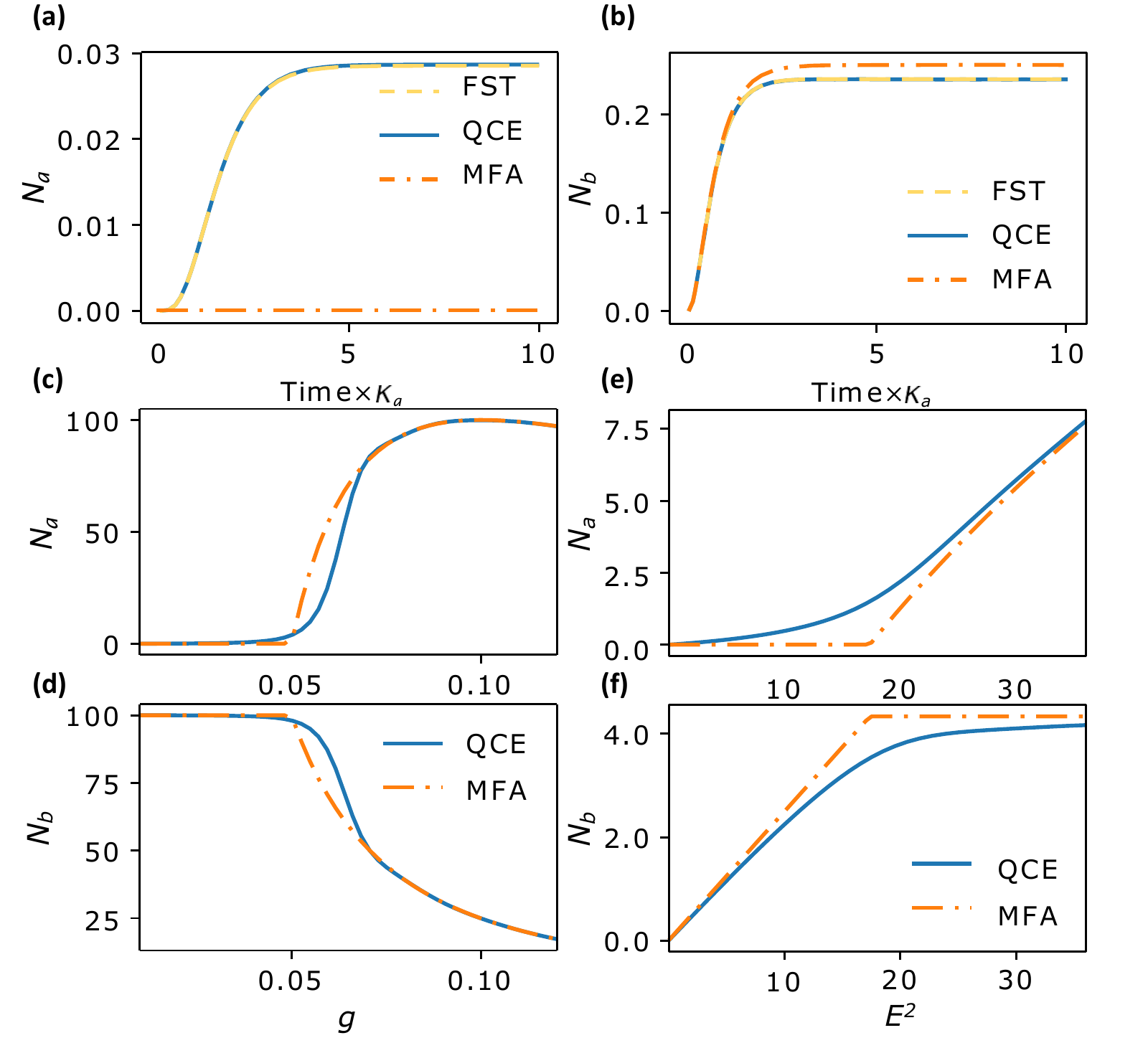}\caption{Quantum-to-classical transition behaviors of optical parametric oscillation.
		(a) and (b) The dynamical evolution of cavity photon numbers, with
		$E=1$ and $g=0.24$. (c)-(d) The steady-state cavity photon number
		against the coupling strength $g$, with $E=20$. (e)-(f) The steady-state
		cavity photon number against the external drive strength $E$, with
		$g=0.24$. In all calculations, $\kappa_{a}=\frac{1}{2}\kappa_{b}=1$  and the evolution duration of the system is set to $T_{ss}=10/\kappa_{a}$.
		Solid lines, dashed lines, and dash-dotted lines are corresponding
		to the results via 4th-order quantum-cluster expansion (QCE), Fock-space
		truncation (FST), and mean-field approximation (MFA), respectively.}
	\label{Fig3}
\end{figure}

Figure~\ref{Fig2} summarizes the results for SHG, under a coherent
drive on mode $a$ as $E\left(a^{\dagger}e^{-i\omega_{d}t}+ae^{i\omega_{d}t}\right)$,
with $E$ is the drive amplitude and $\omega_{d}=\omega_{a}$ is the
frequency of the drive resonant with mode $a$. To validate the QCE, we compare the system evolution modeled by three different approaches.
The master equation provides the most rigorous solution as long as
the FST has the truncated dimension large enough. Due to the limited
computation resource, we take the truncation dimension as $40$ and $20$ for mode $a$ and mode $b$, respectively. For comparison, the QCE is solved with the 4th-order truncation
of clusters. The classical model with MFA $\langle a^{\dagger}b\rangle=\langle a^{\dagger}\rangle\langle b\rangle$
is also evaluated, and we note that the MFA is actually the 1st-order
QCE. For $E=6$ and $g=0.4$, with initial vacuum state, the dynamics
of photon numbers in the two modes show excellent agreement between
QCE and FST. However, the results of MFA deviate from the other two
approaches, indicating the non-negligible quantum correlations between
modes exist for $g/\kappa\sim0.4$ as such effects could not be captured
by the MFA. Additionally, the steady-state populations in two modes
are studies for various $g$ and $E$. Figures~\ref{Fig2}(c) and
(d) show that the MFA and QCE start to deviate when $g$ exceeding
$0.1$, indicating a cross-over from classical to quantum regime. Similar
behavior is shown when increasing $E$ for a fixed $g=0.2$, as shown
in Figs.~\ref{Fig2}(e) and (f). Furthermore, MFA predicts a classical
threshold above which will lead to self-pulsing as marked in the
shadow region. The critical driving strength $E_{c}$ is given
by

\begin{equation}
	E_{c}=\frac{(2\kappa_{a}+\kappa_{b})}{2g}\sqrt{2\kappa_{b}(\kappa_{a}+\kappa_{b})},
\end{equation}
while it can be seen that the result above this threshold does not commit with the ones offered by via QCE as well as FST.

Similar studies on the OPO process is also performed, but with on-resonant coherent
drive applied to mode $b$ as $E\left(b^{\dagger}e^{-i\omega_{d}t}+be^{i\omega_{d}t}\right)$.
As shown in Fig.~\ref{Fig3}, MFA predictions also
significantly deviate from those of FST and QCE. When varying $g$
and $E$, the deviation is only obvious at moderate values around the OPO threshold. As correctly predicted by MFA, degenerate
OPO exhibits threshold driving strength
$E_{c}$ as 
\begin{eqnarray}
	E_{c} & = & \frac{\kappa_{a}\kappa_{b}}{2g},
\end{eqnarray}
similar to a second-order phase transition~\cite{drummond1980non}.
However, after taking the quantum correlation into consideration in the
QCE, the intracavity photon number $\langle a^{\dagger}a\rangle$
and $\langle b^{\dagger}b\rangle$ changes smoothly with the driving
strength $E$, in contrast to the sharp transition predicted by the
MFA [Fig.~\ref{Fig3}(c)]. Due to the quantum fluctuation of the
vacuum field, the spontaneous parametric down-conversion always generates
photons in the fundamental mode $a$, which leads to the non-zero
value of $\langle a^{\dagger}a\rangle$ for $E<E_{c}$. Even though
semi-classical modification can be introduced to MFA to explain the
spontaneous parametric oscillation, it is still difficult to predict the system behavior precisely when the drive amplitude is around the threshold. For both SHG and OPO, MFA is consistent
with other approaches at the weak coupling limit, thus validating
the  coupled-mode theory widely adopted for  nonlinear optical systems
with $g/\kappa\ll1$. For OPO far above threshold ($E$ and $g$ are
large enough), the photon number in the fundamental mode is much larger
than the second-harmonic mode, thus their quantum correlation can
be safely neglected i.e. $\Delta\langle\hat{2}\rangle\approx0$. In
this case, QCE and MFA agree with each other.

\begin{figure}
	\centering{}\includegraphics[width=1\columnwidth]{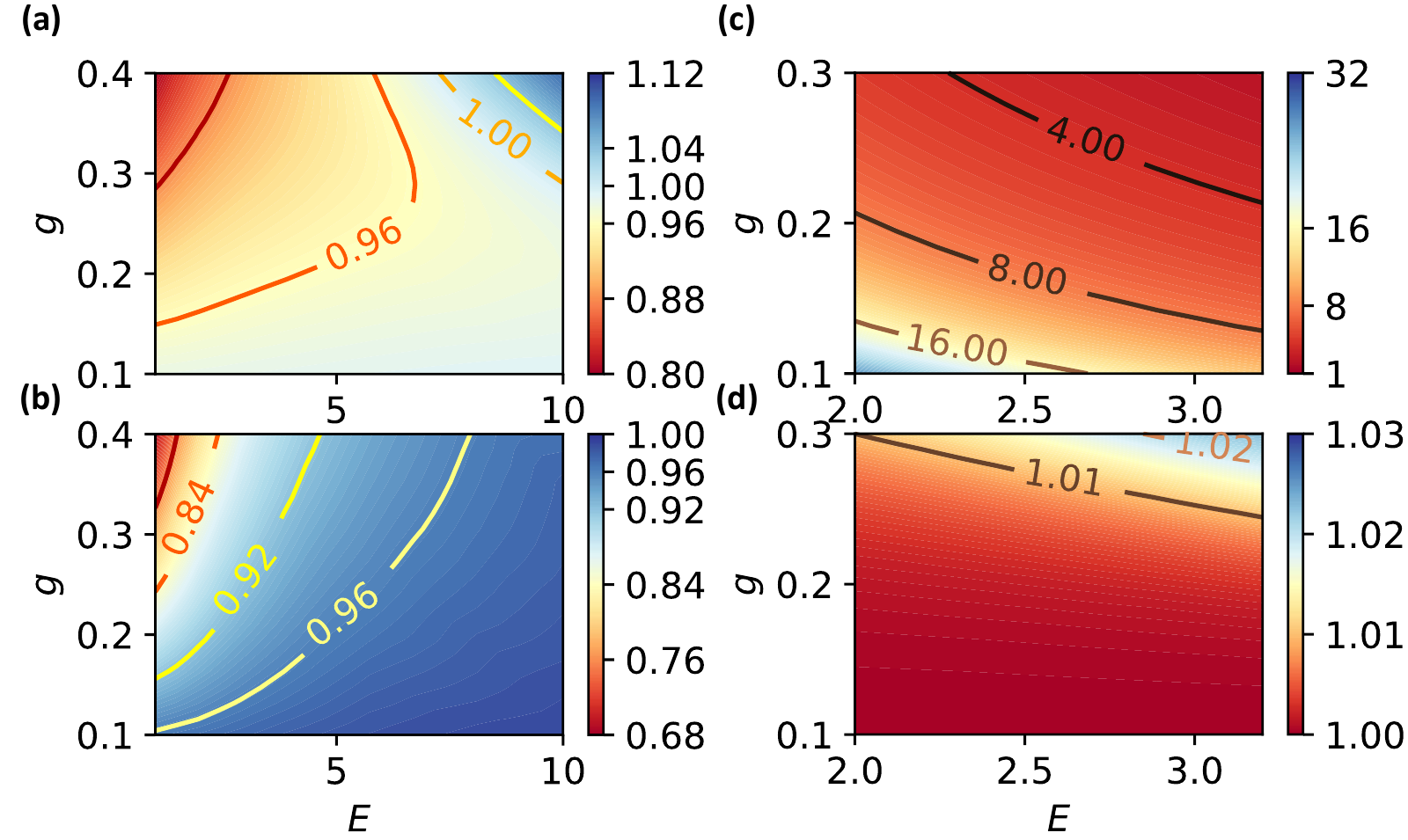}\caption{Quantum second-order correlation function ($g^{(2)}\left(0\right)$)
		of different modes calculated by 6th-order quantum-cluster expansion
		with varying coupling rate $g$ and driving strength $E$. (a,b) The $g^{(2)}\left(0\right)$ for modes $a$ and $b$ , respectively,
		under the SHG drive $\kappa_{a}=\kappa_{b}=1$.
		(c, d) The $g^{(2)}\left(0\right)$ for modes $a$ and $b$,
		respectively, under the OPO drive 
		$\kappa_{a}=\frac{1}{2}\kappa_{b}=1$.For all calculation, the evolution duration of the system is set to $T_{ss}=10/\kappa_{a}$.}
	\label{Fig4}
\end{figure}

Besides offering more accurate prediction of the photon numbers, another
prominent advantage of the QCE is its ability to track the quantum
statistics of the optical fields when the nonlinear optical system
transits from weakly anharmonic to strongly anharmonic regime. In quantum
optics, the second-order self-correlation function
\begin{eqnarray}
	g^{(2)} & = & \frac{\langle O^{\dagger}O^{\dagger}OO\rangle}{\langle O^{\dagger}O\rangle^{2}}
\end{eqnarray}
is often used to quantify the quantum statistics of a mode $O$.
It requires the operator cluster to be truncated at least to the 2nd-order,
whereas MFA could not track the correlation functions since $\langle O^{\dagger}O^{\dagger}OO\rangle=\langle O^{\dagger}O\rangle^{2}$
for $1$st order truncation. In contrary to the coupled-mode equations
by MFA, which can be transformed to $g/\kappa$-invariant form \cite{zheng1995quantum},
the quantum correlation function is coupled with the mean-fields in
QCE, and thus depends on the pump power and $g/\kappa$. It is worth
noting that the QCE also enables the calculation of arbitrary high-order
quantum correlation functions by choosing an appropriate truncation
order.

Figure~\ref{Fig4} shows $g^{(2)}$
as a function of the coupling strength $g$ and the driving strength $E$. For SHG, both the $g^{(2)}$ function of the fundamental and SH mode are smaller than $1$ when $g$ is large and $E$ is small, revealing
the photon anti-bunching due to the significant quantum photon blockade
effect~\cite{KerrBlock,majumdar2013single}. As the driving strength
$E$ becomes large, the $g^{(2)}$ function tends above $1$, which
corresponds to the bunching effect and can be attributed to optical
bistability and bifurcation~\cite{savage1988oscillations,gevorkyan2000bifurcation}.
The OPO case {[}Fig.$\:$\ref{Fig4}(c)-(d){]} also shows pump power-dependent
$g^{(2)}$ function, which indicates the classical-to-quantum transition
of OPO under a strong pump as the value of the $g^{(2)}$ function
diverges from 1.

\begin{figure}[t]
	\begin{centering}
		\includegraphics[width=1\linewidth]{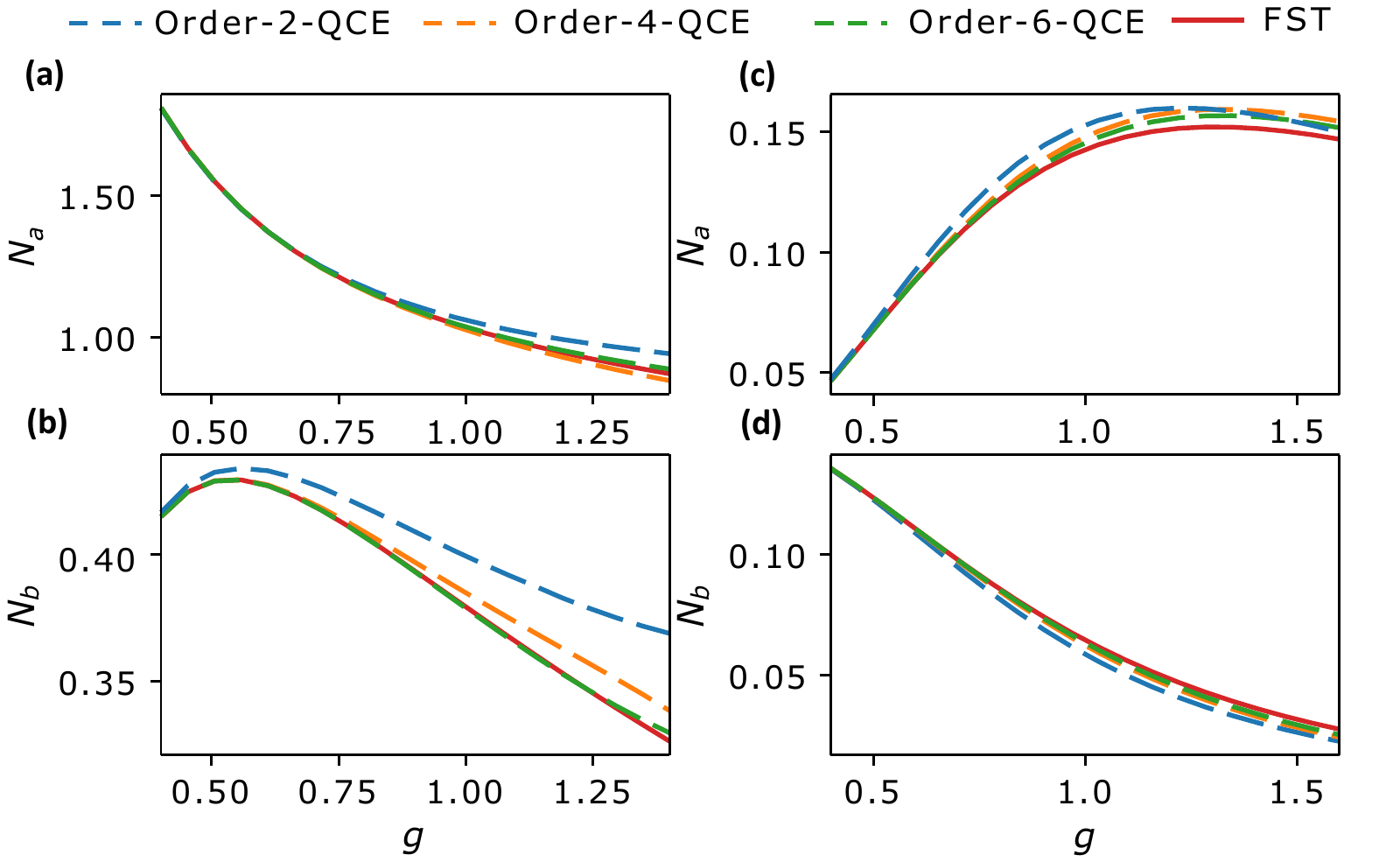}
		\par\end{centering}
	\centering{}\caption{The performance of QEC approach of different orders. (a, b) Steady-state intracavity photon number of modes $a$
		and $b$, under the SHG drive $E=2$ with
		$\kappa_{a}=\kappa_{b}=1$. (c, d) Steady-state
		photon number of modes $a$ and $b$, under the OPO drive intensity $E=0.8$ with $\kappa_{a}=\frac{1}{2}\kappa_{b}=1$. The
		solid lines are the reference results calculated by FST.}
	\label{Fig5}
\end{figure}

\section{Performance analysis}
\begin{figure}[t]
	\centering{}\includegraphics[width=1\columnwidth]{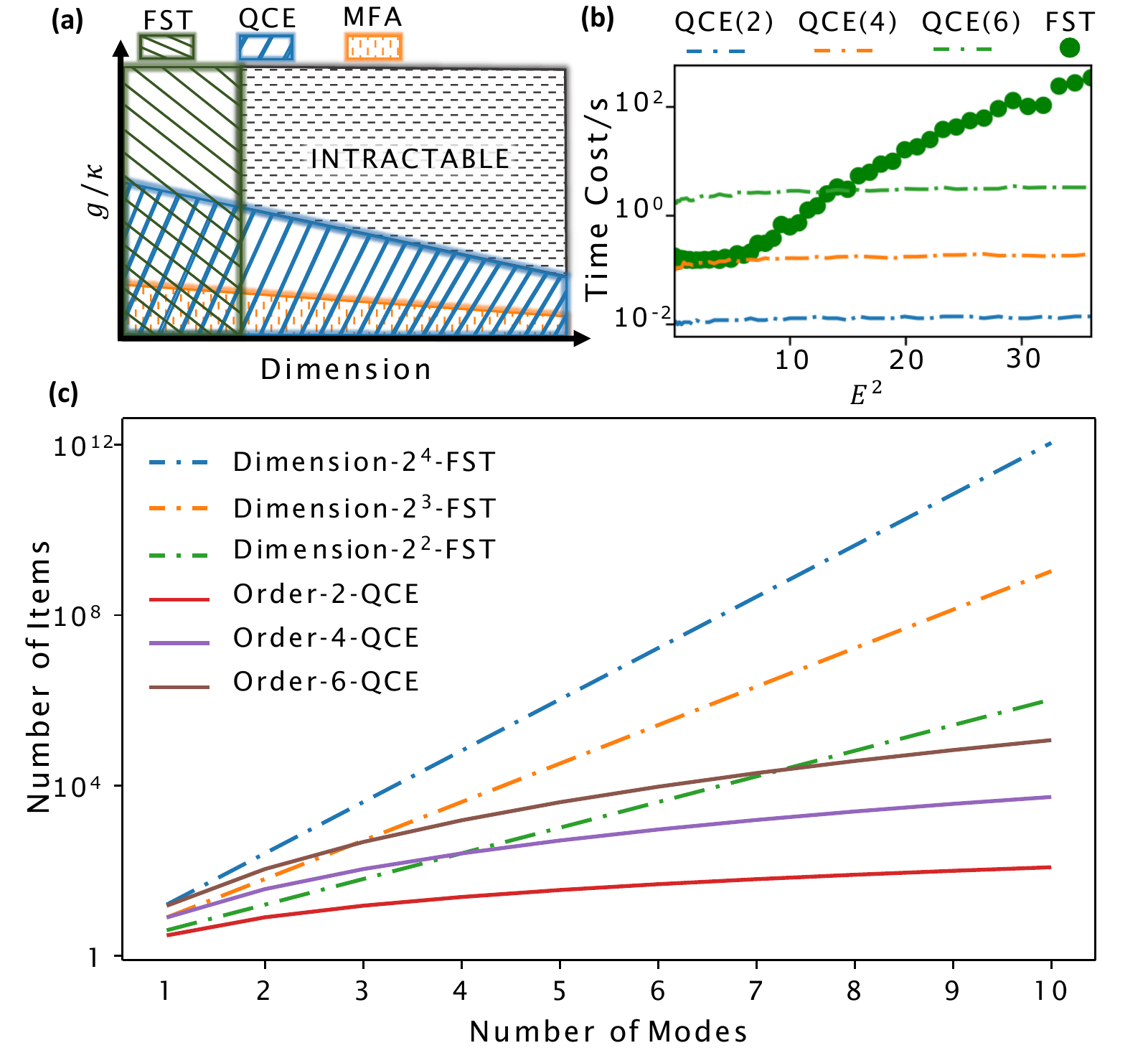}\caption{Time consumption and computational complexity for different approaches. (a) Qualitative illustration of the applicable parameter regions of FST, QCE, and MFA. (b) The computation time of the SHG model via FST compared with that via QEC with different orders on a personal computer. The truncation dimension of FST is $Max(E^2,4)\times \frac{1}{2}Max(E^2,4)$. (c) The computational complexity of multimode systems. Solid lines and dashed-dot lines are the numbers of items of the partial differential equations for QCE with different truncation orders and the FST with different truncation dimensions.}
	\label{Fig6}
\end{figure}
It is of great importance to gain further insights into the performance of QCE as the nonlinear optical system transitions from weakly harmonic ($g/\kappa\ll1$) to strongly anharmonic ($g/\kappa\gg1$) regime. Figure~\ref{Fig5} shows the performances of QCE for SHG and OPO with different QCE truncation orders. The solid line is obtained by solving the master equation in the Fock state basis and is used as a reference. For both SHG and OPO, the deviations of the QCE results become large with the increase of the coupling strength $g$, indicating an increased high-order quantum correlation. It is anticipated that it is more accurate to treat the nonlinear system as multilevel atoms when $g/\kappa\gg1$~\cite{li2020photon}. By expanding the operator clusters to higher-order, the results of QCE, as shown by the dashed curves with orders of $2$, $4$, and $6$, converge to the result of the FST. However, as the order of expansion increases, a much larger number of clusters are involved in the nonlinear coupled equation, and lead to exceptionally high computational complexity, especially for $g\gg\kappa$.

Therefore, the potential advantage of the QCE approach is discussed by comparing its time consumption and computational complexity with other approaches. In a nutshell, Fig.~\ref{Fig6}(a) offers a qualitative illustration of the applicable parameter region for the three methods discussed in this paper. The MFA is efficient and accurate for very weak nonlinearity with $g/\kappa\ll 1$. For the FST, the density matrix contains the full information of the optical fields with a finite Fock space dimension, and can be used for the calculation of the expectation value of operators with arbitrarily high order. Thus, the FST method is particularly powerful for $g/\kappa\gg1$, since only a small amount of photon number states is enough to capture the system's behavior due to the strong anharmonicity, but is limited to very few modes. The QCE is more suitable for nonlinear systems with moderate nonlinearity $g/\kappa\lesssim 1$, and its superiority is particularly significant for strong pump power and large mode number. For example, in Fig.$\:$\ref{Fig2} with $g/\kappa=0.2$, a $4$-th order QCE is enough to predict the photon numbers of both modes with high precision. The star marked near the curve of QCE (Fig.$\:$\ref{Fig2}(d)) presents the result of FST in a dimension of $100\times100$, which costs nearly $12$ hours for $2000$ times Monte Carlo simulations~\cite{zheng1995quantum} by Qutip~\cite{johansson2013qutip}. In contrast, there are only $37$ clusters in the 4-th order QCE used, and the calculate takes only $1$ second on the same computer, showing a speedup over $10^{4}$ times. 

Figure~\ref{Fig6}(b) shows the quantitative result for the time cost to complete a single simulation task via FST and QCE of different orders. It is clear that the time cost for FST increases exponentially with the pump power while the time costs for QCE approaches are almost constants.  Furthermore, the two approaches scale differently with the system dimension, as shown in Fig.$\:$\ref{Fig6}(c). For the open system with $m$ modes, the number of equations to solve with FST is $n_{\mathrm{trunc}}^{m}$ which increases exponentially with $m$ (dashed lines). Thus, the FST approach becomes impractical when the photon number exceeds $1000$ for $m\geq5$, as implied by the quantum supremacy~\cite{preskill2012quantum}. Fortunately, for the QCE, the exponential scaling is reduced down to the polynomial relationship between the number of clusters and the number of modes, as shown by the solid lines in Fig.$\:$\ref{Fig6}(c). To the $2$nd-order expansion, the maximum number of items $f^{(2)}(m)$ tracked in the differential equations is $m^{2}+2m$, which grows quadratically with mode number. For $n$th-order expansion, the number of items is still a polynomial function of mode number with $O(m^{n})$. Even though these clusters couple with each other nonlinearly, the computation complexity still follows a polynomial relationship with the number of modes, thus demonstrating the superiority of QCE for quantum many-body physics in multimode bosonic systems.

\section{Conclusion}

In summary, we use the quantum cluster-expansion approach
to investigate the classical-to-quantum transition of multimode nonlinear optical systems. The $\chi^{(2)}$ nonlinearity is investigated as a signature two-mode system, with pump laser driving either the fundamental mode or second-harmonic mode. We have discussed the system dynamics with various nonlinear coupling strengths to dissipation ratio $g/\kappa$, and different approaches including the MFA, FST, and the QCE are numerically implemented and compared. It is found that the QCE approach could bridge the gap between the MFA and FST, i.e. capture both the classical behaviors and also the quantum correlations between modes, is appropriate for the nonlinear optical system with a large number of modes and also relatively strong excitations with $g/\kappa\lesssim1$. QCE greatly reduces the computational complexity and can be applied to describe a wide range of applications based on bosonic oscillators with moderate anharmonicity, including the rapidly-developing integrated nonlinear photonics$\:$\cite{lu2020toward,chen2021photon,chengya,hao2020second} and the superconducting cavity with kinetic inductance~\cite{wang2019quantum,andersen2020quantum}. The experimental progress raises urgent needs for the theoretical simulation of nonlinear photonic systems with high efficiency.

\smallskip{}

\noindent \textbf{Acknowledgments}\\Y.-X.H. thanks Qianhui Lu and
Mingda Li for their supports and inspiring ideas in numerical calculations.
This work was funded by the National Key R\&D Program (Grant No.~2017YFA0304504), the National Natural Science Foundation of China (Grants No.~11874342, No.~11922411, No.~12061131011, and No.~11904316), Anhui Province Natural Science Foundation (No.~2008085QA34). ML and CLZ were also supported by the Fundamental Research Funds for the Central Universities (Grant Nos. WK2470000031 and WK2030000030), and the State Key Laboratory of Advanced Optical Communication Systems and Networks. The numerical calculations in this paper were partially done on the supercomputing system in the Supercomputing Center of University of Science and Technology of China.

\end{document}